\documentclass[
]{ceurart}

\usepackage{listings}
\usepackage{ulem}
\usepackage{tikz}
\usetikzlibrary{matrix,decorations,arrows,shapes,automata,backgrounds,fit,calc,petri,patterns,positioning}

\newcommand{\nsymbol}{\mathbb{N}}

\begin{document}

\copyrightyear{2024}
\copyrightclause{Copyright for this paper by its authors.
  Use permitted under Creative Commons License Attribution 4.0
  International (CC BY 4.0).}

\conference{}

\title{Petri nets in modelling glucose regulating processes
in the liver}

\author[1]{Kamila Barylska}[%
]

\author[1]{Anna Gogoli{\'n}ska}[%
email=leii@mat.umk.pl,
]

\address[1]{Nicolaus Copernicus University in Toru{\'n}}

\begin{abstract}
Diabetes is a chronic condition, considered one of the civilization diseases, that is characterized by sustained high blood sugar levels. 
There is no doubt that more and more people is going to suffer from diabetes, hence it is crucial to understand better 
its biological foundations. 
The essential processes related to the control of~glucose levels in the blood are: glycolysis (process of breaking down of glucose)
and glucose synthesis, both taking place in the liver.
The~glycolysis occurs during feeding and it is stimulated by insulin.
On the other hand, the glucose synthesis arises during fasting
and it is stimulated by glucagon.
In the paper we present a Petri net model of~glycolysis
and glucose synthesis in the liver. The model is created based
on medical literature. Standard Petri nets techniques are used to
analyse the properties of the model: traps, reachability graphs,
tokens dynamics, deadlocks analysis. The results are described
in the paper.
Our analysis shows that the model captures the interactions between
different enzymes and substances, which is consistent with
the biological processes occurring during fasting and feeding.
The model constitutes the first element of our long-time goal to
create the whole body model of the glucose regulation
in a healthy human and a person with diabetes.
\end{abstract}


\maketitle

\section{Introduction}
Diabetes is a chronic disease that occurs either when the pancreas does not produce enough insulin (a~hormone that regulates blood glucose) or when the body cannot effectively use the insulin  it produces. Diabetes is considered one of the civilization diseases. According to International Diabetes Federation data for 2021 \cite{IDF} and IDF Diabetes Atlas \cite{DA}, one in ten  people in the world, that is, approximately, 537 million adults (20-79 years), are living with diabetes. The total number of people living with diabetes is projected to rise to 643 million by 2030 and 783 million by 2045. It is estimated that 6.7 million adults died from diabetes or its complications in 2021, which means 1 death every 5 seconds. Diabetes was responsible for at least \$966 billion in health expenditure in 2021, which is 9\% of the global total spent on healthcare.

It is therefore not surprising, that in many fields of science intensive work has being undertaken not only to cure the disease, but also, and perhaps above all, to  prevent or delay its future health complications (such as heart disease, chronic kidney disease, nerve damage, and other problems with feet, oral health, vision, hearing, and mental health) or improving the quality of life of patients and their families.

More and more advanced systems are being developed to continuously measure blood glucose levels without the need to puncture the skin (CGM - continuous glucose monitoring \cite{CGM1, CGM2}), the insulin pump industry has been developing rapidly. Closed loop systems (so called artificial pancreas), enabling automatic insulin delivery by the pump, were also created and operate successfully, both as a commercial solution (a.o. MiniMed 780G System~\cite{loop1}, Tandem Tslim Control IQ~\cite{loop2}, CamAPS FX~\cite{loop3}), or developed on a~DIY basis (a.o. AAPS~\cite{loop4}, Loop~\cite{loop5}).
Countless applications are being developed for diabetics and their families, as well as health care professionals, such as bolus calculators, applications for counting food nutritional value, diabetes management, statistics and many others. Intensive work is also underway on the use of artificial intelligence in this field \cite{NN1,NN2,NN3}.
However, we have noticed that most non-medical solutions focus on solving a particular problem without offering a broader view of diabetes. 
On the other hand, medical papers usually focus on one single element of the whole process of glucose regulation for healthy and diabetic people. Therefore, there exist a great need for a more holistic approach, which recently become more and more popular. 

Our long-term goal is to create a simple and intuitive mathematical model representing the changes occurring in the body of a healthy person and a person with diabetes. 
This model should be easily analysable and clear, but at the same time, capable of representing complex processes consisting of interactions between many components. 
In our opinion, Petri nets (PNs) constitute a perfect tool for this purpose. Due to PNs intuitive graphical representation and mathematical properties, the model would be useful for people with and without medical background. 
This could allow for a better understanding of the processes occurring in a human body, predicting new therapeutic targets and designing drug therapies. We are aware, that our goal (modelling the entire process) is ambitious and would not be reached at once.
Hence, our first step, presented below, is to model the glycolysis process. The medical basis of our model is taken from \cite{PAPER}.



Glycolysis 
is the pathway of breakdown of glucose into
pyruvate following glucose uptake by cells. It is an ancient metabolic pathway, meaning that it evolved long ago, and it is found in the great majority of organisms alive today\cite{raven}. The process is highly important in maintaining the homeostasis of glucose levels in the body of a healthy person, especially the glycolysis that occurs in the liver. For that reason  we start
 modelling diabetes 
from this process. Our model allows to scrupulously follow the glycolysis process in a healthy person, as well as to analyse what happens in an organism in which insulin secretion is disturbed or completely absent.
Insulin and glucagon are hormones produced and released by the pancreas to regulate blood sugar levels. Insulin is released by the beta cells of the pancreas and reduces blood sugar levels while glucagon is released by the alpha cells of the pancreas and increases blood sugar levels. The two hormones are responsible for maintaining homeostasis of blood glucose levels. An increase in blood glucose levels triggers the release of insulin which promotes glycolysis, while when glucose levels decrease, glucagon is released to stimulate the opposite process.

In this paper, we present a model of the glycosite 
and the opposite process in the liver designed using a Petri net. We also use standard Petri net analysis tools, such as the reachability graph, to study it. 
In the following section, we recall the basic concepts of Petri nets, 
and in subsequent parts of the paper we introduce 
a Petri net, modelling the processes occurring in the liver
and then conduct an analysis of~it. The paper ends with a summary and future plans.

\section{Preliminaries}

In this section we recall the basic notions concerning Petri Nets and its properties~\cite{petri1,Murata,Reisig,petri2,trap}.

A {\it finite labelled transition system} with initial state is a tuple $TS=(S,\to,T,s_0)$
with nodes $S$ (a~finite set of states),
edge labels $T$ (a~finite set of letters),
edges $\to\,\subseteq(S\times T\times S)$,
and an initial state $s_0\in S$.
A label $t$ is enabled at $s\in S$, denoted by $s[t\rangle$, if $\exists s'\in S\colon(s,t,s')\in\,\to$.
A state $s'$ is reachable from~$s$ through the execution of $\sigma\in T^*$,
denoted by $s[\sigma\rangle s'$, if there is a directed path from~$s$ to~$s'$ whose edges are labelled consecutively by $\sigma$. 
The set of states reachable from $s$ is denoted by $[s\rangle$.
A sequence $\sigma\in T^*$ is allowed, or firable, from a state $s$,
denoted by $s[\sigma\rangle$, if there is some state $s'$ such that $s[\sigma\rangle s'$.

An {\it (initially marked) Petri net} (PN) is denoted as $N=(P,T,F,M_0)$ where $P$ is a finite set of places,
$T$ is a finite set of transitions,
$F$ is the flow function $F\colon((P\times T)\cup(T\times P))\to\nsymbol$ 
specifying the arc weights,
and $M_0$ is the initial marking
(where a marking is a mapping $M\colon P\to\nsymbol$, indicating the number of tokens in each place).
A transition $t\in T$ is enabled at a marking $M$,
denoted by $M[t\rangle$, if $\forall p\in P\colon M(p)\geq F(p,t)$.
The firing of $t$ leads from $M$ to $M'$, denoted by $M[t\rangle M'$,
if $M[t\rangle$ and $M'(p)=M(p)-F(p,t)+F(t,p)$.
This can be extended, as usual, to $M[\sigma\rangle M'$ for sequences $\sigma\in T^*$,
and $[M\rangle$ denotes the set of markings reachable from $M$.
We call a marking $M$ {\it deadlock} if it does not enable any transition, i.e. for every $t\in T$ we have $\exists p\in P\colon M(p)< F(p,t)$.
The reachability graph $RG(N)$ of a bounded (such that the number of tokens
in each place does not exceed a certain finite number) Petri net $N$
is the labelled transition system with the set of vertices $[M_0\rangle$,
initial state $M_0$, label set $T$,
and set of edges $\{(M,t,M')\mid M,M'\in[M_0\rangle\land M[t\rangle M'\}$.

\begin{figure}[htb]
\begin{center}
\hbox{}
\begin{tikzpicture}
\node[draw,minimum size=0.5cm](a)at(0,1.5){$a$};
\node[draw,minimum size=0.5cm](b)at(2,1.5){$b$};
\node[draw,minimum size=0.5cm](c)at(3,1.5){$c$};
\node[circle,draw,thick,minimum size=0.4cm](s0)at(1,3){};\filldraw[black](1,3)circle(2pt);\draw(0.5,3)node{$p$};
\node[circle,draw,minimum size=0.4cm](s1)at(1,2){};\filldraw[black](1,2)circle(2pt);
\node[circle,draw,minimum size=0.4cm](s2)at(1,1){};
\node[circle,draw,minimum size=0.4cm](s3)at(1,0){};
\draw[-latex](s0)--(a);
\draw[-latex](s0)--(b);
\draw[-latex](s1)--(a);
\draw[-latex](b)--(s1);
\draw[-latex](a)--(s2);
\draw[-latex](s2)--(b);
\draw[-latex](a)--(s3);
\draw[-latex](b)--(s3);
\draw[-latex](s3)to[out=0,in=270](c);
\draw[-latex](c)to[out=90,in=0](s0);
\end{tikzpicture}\hspace*{1cm}
\raisebox{1cm}{\begin{tikzpicture}[scale=0.6]
\node[circle,fill=black!100,inner sep=0.05cm](0)at(0,2)[label=above left:$M_0$]{};
\node[circle,fill=black!100,inner sep=0.05cm](1)at(2,2)[label=above:]{};
\node[circle,fill=black!100,inner sep=0.05cm](2)at(2,0)[label=above:]{};
\node[circle,fill=black!100,inner sep=0.05cm](3)at(0,0)[label=above:]{};
\draw[-triangle 45](0)to[]node[above,swap,inner sep=2pt]{$a$}(1);
\draw[-triangle 45](1)to[]node[right,swap,inner sep=2pt]{$c$}(2);
\draw[-triangle 45](2)to[]node[above,swap,inner sep=2pt]{$b$}(3);
\draw[-triangle 45](3)to[]node[left,swap,inner sep=2pt]{$c$}(0);
\end{tikzpicture}}
\label{netgraph}
\caption{A Petri net and its reachability graph.}
 \vspace*{-0.5cm}
\end{center}
\end{figure}
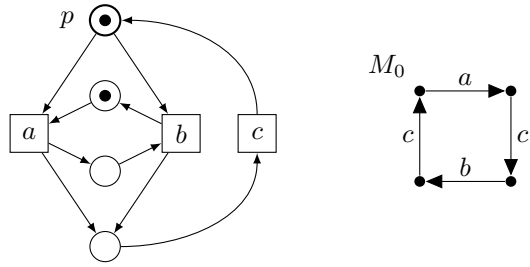

Note that the reachability graph of a bounded Petri net captures the exact information about the reachable markings of the net, and therefore reflects the entire behaviour of a given net. Figure~\ref{netgraph} depicts a Petri net, together with its reachability graph. Reachability graphs of real biological systems are usually quite large and therefore difficult to analyse. To deal with this inconvenience, we use reduced reachability graphs, called {\it stubborn reduced reachability graphs}, created on the basis of partial order reduction techniques, where not all interleaving sequences of
concurrent behaviour are considered. 
As~a~result of the reduction only a subset of the complete reachability graph is constructed, nevertheless it still allows the discussion of certain properties, in particular: it preserves all deadlock states and the whole cyclic behaviour.
\\
The reduction of a reachability graph to
 a stubborn reduced reachability graph proceeds as follows:
\begin{enumerate}
\item For a given marking, determine a set of "independent" transitions (called {\it stubborn set}), such that their behaviour cannot be influenced by any transitions from the complement 
of this set (i.e.~\textit{excluded} transitions). 
Additionally, the following conditions must hold: any sequence of excluded transitions cannot enable or disable an included transition (hence their firing can be postponed) and the set contains at least one enabled transition.
\item Compute a \textit{stubborn reduced reachability graph}, using a variation of a standard algorithm: 
at~each marking (node), instead of firing all enabled transitions, 
 only transitions of a stubborn set are fired.
\end{enumerate}
The notion of stubborn sets capture the lack of interaction between transitions,  and such excluded transitions
 may not be interesting from our point of view (for instance in case of  biological systems). Executions of transitions from outside a stubborn set could be postponed because it does not affect the merits of the system's behaviour\footnote{Due to lack of space, we do not provide detailed definitions and 
properties here, interested readers are referred to the literature (a.o. \cite{valmari1}, \cite{valmari2}, \cite{heiner}, \cite{charlie}).}.

Let $x \in P \cup T$, $^\bullet x = \{y\in (P \cup T) \mid (y,x) \in F\}$ and
$ x ^\bullet = \{y\in (P \cup T) \mid (x, y) \in F\}$. We extend this notation to sets: 
for $S \subseteq P \cup T$, we have $^\bullet S = \bigcup_{x \in S} {^\bullet x}$ and
$S^\bullet = \bigcup_{x \in S} x^\bullet$. 
Let $S$ be a~non-empty subset of places. We call $S \subseteq P$
a \textit{trap} if $S ^\bullet \subseteq ^\bullet S$. The following property follows directly from the definition: once a trap is marked under some marking, it is always marked at the subsequent markings reachable from this one.

\section{Model}
\label{model}

Our Petri Net model, presented in Figure~\ref{fig2}, represents the process of glycolysis and the opposite activity in a liver described in~\cite{PAPER}. It contains 17 places and 13 transitions. 

Homeostasis is the state of steady chemical conditions maintained by organism. In our paper we focuse on the glucose homeostasis. 
There are three basic organs of the human body that control the level of glucose: the pancreas, the liver and the fat tissue. The goal of our PN model is to represent the processes occurring in the liver that controll the level of glucose. Two basic processes are present, namely: glycolysis and synthesis of glucose. 

The glycolysis in the liver is stimulated by the presence of insulin and glucose, which physiologically
is associated with feeding. Those substances, in the PN model, are represented by places: \textcolor{red}{\textit{Insulin}} and \textcolor{red}{\textit{Glucose}} respectively. 
The presence of insulin activates various protein phosphatases 
(place~\textcolor{red}{\textit{PP}}). It results in 
an increase in the expression of glucokinase (place~\textcolor{red}{\textit{GK}})
and leads to an increasing the kinase activity of complex 6PFK2/FBPase2 (place~\textcolor{red}{\textit{Kin}})
and a decreasing the phosphatase activity of 6PFK2/FBPase2 (place~\textcolor{red}{\textit{Pho}}).
As~a~result of kinase activity of 6PFK2/FBPase2, the level of fructose-2,6-bisphosphate 
(F2, 6P2) is  increasing (place~\textcolor{red}{\textit{F26P2}}), leading to an elevation of
6-phosphofructo-1-kinase (6PFK1) activity (place~\textcolor{red}{\textit{a6PFK1}}) and a
reduction of fructose-1,6-bisphosphatase (FBPase) activity (place~\textcolor{red}{\textit{nFBPase}}).
All~those changes cause an increasement in glycolysis rates. 
The glycolysis is represented by 
the part of the net (places and transitions) leading from place \textit{Glucose}, through place~\textit{F6P} to place~\textcolor{red}{\textit{Pyruvate}},
which represents the final product of the glycolysis.
When the feeding is long time gone and the fasting occurs the levels of glucose and insulin in the blood are very low. It provokes a production of glucagon by the pancreas.
The presence of this enzyme is represented by place~\textcolor{red}{\textit{Glucagon}}. By the signal cascade, 
(places~\textcolor{red}{\textit{cAMP}} and ~\textcolor{red}{\textit{PKA}}), glucagon alters the level of glucokinase and of phosphorylation
of complex 6PFK2/FBPase2. It results in an increase of the phosphatase activity of 6PFK2/FBPase2 
(place~{\textit{Pho}) and a decrease of the kinase activity of 6PFK2/FBPase2 (place~\textit{Kin}).
The phosphatase activity of 6PFK2/FBPase2 causes a decrease of F2,6P2 level
(place ~\textcolor{red}{\textit{mF26p2}}). It leads to a reduction of 6PFK1 activity (place~\textcolor{red}{\textit{n6PFK1}}) 
and an elevation of FBPase activity (place~\textcolor{red}{\textit{aFBPase}}). Those enzymes inhibit
the glycolysis process and stimulates the synthesis of glucose from pyruvate.

\begin{figure}[ht]
\begin{center}
\begin{tikzpicture}[x=1pt,y=-1pt, scale=0.52]

\definecolor{BLACK}{RGB}{0,0,0}
\definecolor{WHITE}{RGB}{255,255,255}
\draw[BLACK, solid, line join=round, line cap=round, line width=1, fill=WHITE]
	(130,110) ellipse[x radius=10, y radius=10];
\draw[BLACK]
	(140,110) node[rotate=0, font=\ttfamily\normalsize, BLACK, right=-.25]
	{\textcolor{red}{Glucagon}};
\draw[BLACK, solid, line join=round, line cap=round, line width=1, fill=WHITE]
	(522,110) ellipse[x radius=10, y radius=10];
\draw[BLACK]
	(536,110) node[rotate=0, font=\ttfamily\normalsize, BLACK, right=-.25]
	{\textcolor{red}{Glucose}};
\draw[BLACK, solid, line join=round, line cap=round, line width=1, fill=BLACK]
	(518,110) ellipse[x radius=3, y radius=3];
\draw[BLACK, solid, line join=round, line cap=round, line width=1, fill=BLACK]
	(526,110) ellipse[x radius=3, y radius=3];
\draw[BLACK, solid, line join=round, line cap=round, line width=1, fill=WHITE]
	(880,110) ellipse[x radius=10, y radius=10];
\draw[BLACK]
	(771,110) node[rotate=0, font=\ttfamily\normalsize, BLACK, right=-.25]
	{\textcolor{red}{Inslulin}};
\draw[BLACK, solid, line join=round, line cap=round, line width=1, fill=BLACK]
	(880,110) ellipse[x radius=3, y radius=3];
\draw[BLACK, solid, line join=round, line cap=round, line width=1, fill=WHITE]
	(130,300) ellipse[x radius=10, y radius=10];
\draw[BLACK]
	(140,300) node[rotate=0, font=\ttfamily\normalsize, BLACK, right=-.25]
	{\textcolor{red}{cAMP}};
\draw[BLACK, solid, line join=round, line cap=round, line width=1, fill=WHITE]
	(130,470) ellipse[x radius=10, y radius=10];
\draw[BLACK]
	(140,470) node[rotate=0, font=\ttfamily\normalsize, BLACK, right=-.25]
	{\textcolor{red}{PKA}};
\draw[BLACK, solid, line join=round, line cap=round, line width=1, fill=WHITE]
	(400,670) ellipse[x radius=10, y radius=10];
\draw[BLACK]
	(408,680) node[rotate=0, font=\ttfamily\normalsize, BLACK, right=-.25]
	{\textcolor{red}{Pho}};
\draw[BLACK, solid, line join=round, line cap=round, line width=1, fill=BLACK]
	(400,670) ellipse[x radius=3, y radius=3];
\draw[BLACK, solid, line join=round, line cap=round, line width=1, fill=WHITE]
	(860,670) ellipse[x radius=10, y radius=10];
\draw[BLACK]
	(865,680) node[rotate=0, font=\ttfamily\normalsize, BLACK, right=-.25]
	{\textcolor{red}{Kin}};
\draw[BLACK, solid, line join=round, line cap=round, line width=1, fill=BLACK]
	(861,670) ellipse[x radius=3, y radius=3];
\draw[BLACK, solid, line join=round, line cap=round, line width=1, fill=WHITE]
	(880,290) ellipse[x radius=10, y radius=10];
\draw[BLACK]
	(838,290) node[rotate=0, font=\ttfamily\normalsize, BLACK, right=-.25]
	{\textcolor{red}{PP}};
\draw[BLACK, solid, line join=round, line cap=round, line width=1, fill=WHITE]
	(730,200) ellipse[x radius=10, y radius=10];
\draw[BLACK]
	(742,200) node[rotate=0, font=\ttfamily\normalsize, BLACK, right=-.25]
	{\textcolor{red}{GK}};
\draw[BLACK, solid, line join=round, line cap=round, line width=1, fill=BLACK]
	(730,200) ellipse[x radius=3, y radius=3];
\draw[BLACK, solid, line join=round, line cap=round, line width=1, fill=WHITE]
	(524,273) ellipse[x radius=10, y radius=10];
\draw[BLACK]
	(536,267) node[rotate=0, font=\ttfamily\normalsize, BLACK, right=-.25]
	{\textcolor{red}{F6P}};
\draw[BLACK, solid, line join=round, line cap=round, line width=1, fill=WHITE]
	(531,519) ellipse[x radius=10, y radius=10];
\draw[BLACK]
	(500,542) node[rotate=0, font=\ttfamily\normalsize, BLACK, right=-.25]
	{\textcolor{red}{F26P2}};
\draw[BLACK, solid, line join=round, line cap=round, line width=1, fill=BLACK]
	(530,520) ellipse[x radius=3, y radius=3];
\draw[BLACK, solid, line join=round, line cap=round, line width=1, fill=WHITE]
	(270,390) ellipse[x radius=10, y radius=10];
\draw[BLACK]
	(215,411) node[rotate=0, font=\ttfamily\normalsize, BLACK, right=-.25]
	{\textcolor{red}{a6PFK1}};
\draw[BLACK, solid, line join=round, line cap=round, line width=1, fill=BLACK]
	(270,390) ellipse[x radius=3, y radius=3];
\draw[BLACK, solid, line join=round, line cap=round, line width=1, fill=WHITE]
	(252,503) ellipse[x radius=10, y radius=10];
\draw[BLACK]
	(220,522) node[rotate=0, font=\ttfamily\normalsize, BLACK, right=-.25]
	{\textcolor{red}{n6PFK1}};
\draw[BLACK, solid, line join=round, line cap=round, line width=1, fill=BLACK]
	(252,503) ellipse[x radius=3, y radius=3];
\draw[BLACK, solid, line join=round, line cap=round, line width=1, fill=WHITE]
	(750,360) ellipse[x radius=10, y radius=10];
\draw[BLACK]
	(740,380) node[rotate=0, font=\ttfamily\normalsize, BLACK, right=-.25]
	{\textcolor{red}{aFBPase}};
\draw[BLACK, solid, line join=round, line cap=round, line width=1, fill=BLACK]
	(750,360) ellipse[x radius=3, y radius=3];
\draw[BLACK, solid, line join=round, line cap=round, line width=1, fill=WHITE]
	(760,430) ellipse[x radius=10, y radius=10];
\draw[BLACK]
	(740,451) node[rotate=0, font=\ttfamily\normalsize, BLACK, right=-.25]
	{\textcolor{red}{nFBPase}};
\draw[BLACK, solid, line join=round, line cap=round, line width=1, fill=BLACK]
	(760,430) ellipse[x radius=3, y radius=3];
\draw[BLACK, solid, line join=round, line cap=round, line width=1, fill=WHITE]
	(490,370) ellipse[x radius=10, y radius=10];
\draw[BLACK]
	(450,390) node[rotate=0, font=\ttfamily\normalsize, BLACK, right=-.25]
	{\textcolor{red}{Pyruvate}};
\draw[BLACK, solid, line join=round, line cap=round, line width=1, fill=WHITE]
	(540,660) ellipse[x radius=10, y radius=10];
\draw[BLACK]
	(537,680) node[rotate=0, font=\ttfamily\normalsize, BLACK, right=-.25]
	{\textcolor{red}{mF26P2}};
\draw[BLACK, solid, line join=round, line cap=round, line width=1, fill=BLACK]
	(540,660) ellipse[x radius=3, y radius=3];
\draw[BLACK, solid, line join=round, line cap=round, line width=1, fill=WHITE]
	(120,210) rectangle +(20,20);
\draw[BLACK]
	(120,220) node[rotate=0, font=\ttfamily\normalsize, BLACK, right=-.25]
	{\textcolor{blue}{\scriptsize{$t_0$}}};
\draw[BLACK, solid, line join=round, line cap=round, line width=1, fill=WHITE]
	(120,390) rectangle +(20,20);
\draw[BLACK]
	(120,399.5) node[rotate=0, font=\ttfamily\normalsize, BLACK, right=-.25]
	{\textcolor{blue}{\scriptsize{$t_1$}}};
\draw[BLACK, solid, line join=round, line cap=round, line width=1, fill=WHITE]
	(200,580) rectangle +(20,20);
\draw[BLACK]
	(197,592) node[rotate=0, font=\ttfamily\normalsize, BLACK, right=-.25]
	{\textcolor{blue}{\scriptsize{$t_2$}}};
\draw[BLACK, solid, line join=round, line cap=round, line width=1, fill=WHITE]
	(870,190) rectangle +(20,20);
\draw[BLACK]
	(870,200) node[rotate=0, font=\ttfamily\normalsize, BLACK, right=-.25]
	{\textcolor{blue}{\scriptsize{$t_3$}}};
\draw[BLACK, solid, line join=round, line cap=round, line width=1, fill=WHITE]
	(870,460) rectangle +(20,20);
\draw[BLACK]
	(870,470) node[rotate=0, font=\ttfamily\normalsize, BLACK, right=-.25]
	{\textcolor{blue}{\scriptsize{$t_4$}}};
\draw[BLACK, solid, line join=round, line cap=round, line width=1, fill=WHITE]
	(570,160) rectangle +(20,20);
\draw[BLACK]
	(569,168) node[rotate=0, font=\ttfamily\normalsize, BLACK, right=-.25]
	{\textcolor{blue}{\scriptsize{$t_6$}}};
\draw[BLACK, solid, line join=round, line cap=round, line width=1, fill=WHITE]
	(459,160) rectangle +(20,20);
\draw[BLACK]
	(458,168) node[rotate=0, font=\ttfamily\normalsize, BLACK, right=-.25]
	{\textcolor{blue}{\scriptsize{$t_5$}}};
\draw[BLACK, solid, line join=round, line cap=round, line width=1, fill=WHITE]
	(350,580) rectangle +(20,20);
\draw[BLACK]
	(349,590) node[rotate=0, font=\ttfamily\normalsize, BLACK, right=-.25]
	{\textcolor{blue}{\scriptsize{$t_7$}}};
\draw[BLACK, solid, line join=round, line cap=round, line width=1, fill=WHITE]
	(740,600) rectangle +(20,20);
\draw[BLACK]
	(740,612) node[rotate=0, font=\ttfamily\normalsize, BLACK, right=-.25]
	{\textcolor{blue}{\scriptsize{$t_8$}}};
\draw[BLACK, solid, line join=round, line cap=round, line width=1, fill=WHITE]
	(330,450) rectangle +(20,20);
\draw[BLACK]
	(327,460) node[rotate=0, font=\ttfamily\normalsize, BLACK, right=-.25]
	{\textcolor{blue}{\scriptsize{$t_9$}}};
\draw[BLACK, solid, line join=round, line cap=round, line width=1, fill=WHITE]
	(670,460) rectangle +(21,21);
\draw[BLACK]
	(665,471) node[rotate=0, font=\ttfamily\normalsize, BLACK, right=-.25]
	{\textcolor{blue}{\scriptsize{$t_{10}$}}};
\draw[BLACK, solid, line join=round, line cap=round, line width=1, fill=WHITE]
	(420,310) rectangle +(21,21);
\draw[BLACK]
	(415,321) node[rotate=0, font=\ttfamily\normalsize, BLACK, right=-.25]
	{\textcolor{blue}{\scriptsize{$t_{11}$}}};
\draw[BLACK, solid, line join=round, line cap=round, line width=1, fill=WHITE]
	(650,310) rectangle +(21,21);
\draw[BLACK]
	(645,321) node[rotate=0, font=\ttfamily\normalsize, BLACK, right=-.25]
	{\textcolor{blue}{\scriptsize{$t_{12}$}}};
\draw[BLACK, solid, line join=round, line cap=round, line width=1]
	(130,120) -- (130,210);
\draw[BLACK, solid, line join=round, line cap=round, line width=1, fill=BLACK]
	(130,210) -- (127,200) -- (133,200) -- (130,210) -- cycle;
\draw[BLACK, solid, line join=round, line cap=round, line width=1]
	(130,230) -- (130,290);
\draw[BLACK, solid, line join=round, line cap=round, line width=1, fill=BLACK]
	(127,280) -- (133,280) -- (130,290) -- (127,280) -- cycle;
\draw[BLACK, solid, line join=round, line cap=round, line width=1]
	(130,310) -- (130,390);
\draw[BLACK, solid, line join=round, line cap=round, line width=1, fill=BLACK]
	(127,380) -- (133,380) -- (130,390) -- (127,380) -- cycle;
\draw[BLACK, solid, line join=round, line cap=round, line width=1]
	(130,410) -- (130,460);
\draw[BLACK, solid, line join=round, line cap=round, line width=1, fill=BLACK]
	(127,450) -- (133,450) -- (130,460) -- (127,450) -- cycle;
\draw[BLACK, solid, line join=round, line cap=round, line width=1]
	(130,481) -- (203,580);
\draw[BLACK, solid, line join=round, line cap=round, line width=1, fill=BLACK]
	(203,580) -- (194,574) -- (199,570) -- (203,580) -- cycle;
\draw[BLACK, solid, line join=round, line cap=round, line width=1]
	(220,594) -- (390,666);
\draw[BLACK, solid, line join=round, line cap=round, line width=1, fill=BLACK]
	(390,666) -- (379,665) -- (382,659) -- (390,666) -- cycle;
\draw[BLACK, solid, line join=round, line cap=round, line width=1]
	(850,669) -- (220,591);
\draw[BLACK, solid, line join=round, line cap=round, line width=1, fill=BLACK]
	(220,591) -- (230,589) -- (230,596) -- (220,591) -- cycle;
\draw[BLACK, solid, line join=round, line cap=round, line width=1]
	(880,120) -- (880,190);
\draw[BLACK, solid, line join=round, line cap=round, line width=1, fill=BLACK]
	(880,190) -- (877,180) -- (883,180) -- (880,190) -- cycle;
\draw[BLACK, solid, line join=round, line cap=round, line width=1]
	(880,210) -- (880,280);
\draw[BLACK, solid, line join=round, line cap=round, line width=1, fill=BLACK]
	(877,270) -- (883,270) -- (880,280) -- (877,270) -- cycle;
\draw[BLACK, solid, line join=round, line cap=round, line width=1]
	(880,300) -- (880,460);
\draw[BLACK, solid, line join=round, line cap=round, line width=1, fill=BLACK]
	(877,450) -- (883,450) -- (880,460) -- (877,450) -- cycle;
\draw[BLACK, solid, line join=round, line cap=round, line width=1]
	(880,480) -- (862,660);
\draw[BLACK, solid, line join=round, line cap=round, line width=1, fill=BLACK]
	(862,660) -- (860,650) -- (866,651) -- (862,660) -- cycle;
\draw[BLACK, solid, line join=round, line cap=round, line width=1]
	(410,666) -- (870,470);
\draw[BLACK, solid, line join=round, line cap=round, line width=1, fill=BLACK]
	(870,470) -- (863,477) -- (859,471) -- (870,470) -- cycle;
\draw[BLACK, solid, line join=round, line cap=round, line width=1]
	(531,112) -- (571,160);
\draw[BLACK, solid, line join=round, line cap=round, line width=1, fill=BLACK]
	(571,160) -- (562,155) -- (568,150) -- (571,160) -- cycle;
\draw[BLACK, solid, line join=round, line cap=round, line width=1]
	(850,665) -- (760,615);
\draw[BLACK, solid, line join=round, line cap=round, line width=1, fill=BLACK]
	(760,615) -- (770,617) -- (767,623) -- (760,615) -- cycle;
\draw[BLACK, solid, line join=round, line cap=round, line width=1]
	(365,600) -- (395,660);
\draw[BLACK, solid, line join=round, line cap=round, line width=1, fill=BLACK]
	(395,660) -- (388,653) -- (394,650) -- (395,660) -- cycle;
\draw[BLACK, solid, line join=round, line cap=round, line width=1]
	(403,660) -- (420,610) -- (370,593);
\draw[BLACK, solid, line join=round, line cap=round, line width=1, fill=BLACK]
	(370,593) -- (381,593) -- (378,600) -- (370,593) -- cycle;
\draw[BLACK, solid, line join=round, line cap=round, line width=1]
	(760,608) -- (840,590) -- (858,660);
\draw[BLACK, solid, line join=round, line cap=round, line width=1, fill=BLACK]
	(858,660) -- (852,651) -- (858,649) -- (858,660) -- cycle;
\draw[BLACK, solid, line join=round, line cap=round, line width=1]
	(477,160) -- (514,112);
\draw[BLACK, solid, line join=round, line cap=round, line width=1, fill=BLACK]
	(514,112) -- (511,122) -- (505,118) -- (514,112) -- cycle;
\draw[BLACK, solid, line join=round, line cap=round, line width=1]
	(520,524) -- (370,586);
\draw[BLACK, solid, line join=round, line cap=round, line width=1, fill=BLACK]
	(370,586) -- (378,579) -- (381,585) -- (370,586) -- cycle;
\draw[BLACK, solid, line join=round, line cap=round, line width=1]
	(740,606) -- (540,524);
\draw[BLACK, solid, line join=round, line cap=round, line width=1, fill=BLACK]
	(540,524) -- (551,525) -- (548,531) -- (540,524) -- cycle;
\draw[BLACK, solid, line join=round, line cap=round, line width=1]
	(280,400) -- (330,450);
\draw[BLACK, solid, line join=round, line cap=round, line width=1, fill=BLACK]
	(330,450) -- (321,445) -- (325,441) -- (330,450) -- cycle;
\draw[BLACK, solid, line join=round, line cap=round, line width=1]
	(330,465) -- (262,498);
\draw[BLACK, solid, line join=round, line cap=round, line width=1, fill=BLACK]
	(262,498) -- (270,491) -- (272,497) -- (262,498) -- cycle;
\draw[BLACK, solid, line join=round, line cap=round, line width=1]
	(750,431) -- (350,459);
\draw[BLACK, solid, line join=round, line cap=round, line width=1, fill=BLACK]
	(350,459) -- (360,455) -- (360,462) -- (350,459) -- cycle;
\draw[BLACK, solid, line join=round, line cap=round, line width=1]
	(350,458) -- (740,362);
\draw[BLACK, solid, line join=round, line cap=round, line width=1, fill=BLACK]
	(740,362) -- (731,368) -- (729,362) -- (740,362) -- cycle;
\draw[BLACK, solid, line join=round, line cap=round, line width=1]
	(670,468) -- (280,392);
\draw[BLACK, solid, line join=round, line cap=round, line width=1, fill=BLACK]
	(280,392) -- (290,391) -- (289,397) -- (280,392) -- cycle;
\draw[BLACK, solid, line join=round, line cap=round, line width=1]
	(262,502) -- (670,471);
\draw[BLACK, solid, line join=round, line cap=round, line width=1, fill=BLACK]
	(670,471) -- (660,475) -- (660,468) -- (670,471) -- cycle;
\draw[BLACK, solid, line join=round, line cap=round, line width=1]
	(744,370) -- (686,460);
\draw[BLACK, solid, line join=round, line cap=round, line width=1, fill=BLACK]
	(686,460) -- (689,450) -- (695,453) -- (686,460) -- cycle;
\draw[BLACK, solid, line join=round, line cap=round, line width=1]
	(690,465) -- (750,435);
\draw[BLACK, solid, line join=round, line cap=round, line width=1, fill=BLACK]
	(750,435) -- (743,442) -- (740,436) -- (750,435) -- cycle;
\draw[BLACK, solid, line join=round, line cap=round, line width=1]
	(514,278) -- (440,315);
\draw[BLACK, solid, line join=round, line cap=round, line width=1, fill=BLACK]
	(440,315) -- (447,308) -- (450,314) -- (440,315) -- cycle;
\draw[BLACK, solid, line join=round, line cap=round, line width=1]
	(440,328) -- (480,362);
\draw[BLACK, solid, line join=round, line cap=round, line width=1, fill=BLACK]
	(480,362) -- (470,358) -- (474,353) -- (480,362) -- cycle;
\draw[BLACK, solid, line join=round, line cap=round, line width=1]
	(280,386) -- (420,324);
\draw[BLACK, solid, line join=round, line cap=round, line width=1, fill=BLACK]
	(420,324) -- (412,331) -- (410,325) -- (420,324) -- cycle;
\draw[BLACK, solid, line join=round, line cap=round, line width=1]
	(420,319) -- (320,310) -- (276,380);
\draw[BLACK, solid, line join=round, line cap=round, line width=1, fill=BLACK]
	(276,380) -- (279,370) -- (284,373) -- (276,380) -- cycle;
\draw[BLACK, solid, line join=round, line cap=round, line width=1]
	(750,350) -- (750,300) -- (670,318);
\draw[BLACK, solid, line join=round, line cap=round, line width=1, fill=BLACK]
	(670,318) -- (679,312) -- (680,319) -- (670,318) -- cycle;
\draw[BLACK, solid, line join=round, line cap=round, line width=1]
	(670,324) -- (740,356);
\draw[BLACK, solid, line join=round, line cap=round, line width=1, fill=BLACK]
	(740,356) -- (730,355) -- (732,348) -- (740,356) -- cycle;
\draw[BLACK, solid, line join=round, line cap=round, line width=1]
	(500,367) -- (650,323);
\draw[BLACK, solid, line join=round, line cap=round, line width=1, fill=BLACK]
	(650,323) -- (641,329) -- (639,323) -- (650,323) -- cycle;
\draw[BLACK, solid, line join=round, line cap=round, line width=1]
	(650,317) -- (534,276);
\draw[BLACK, solid, line join=round, line cap=round, line width=1, fill=BLACK]
	(534,276) -- (545,277) -- (542,283) -- (534,276) -- cycle;
\draw[BLACK, solid, line join=round, line cap=round, line width=1]
	(550,658) -- (740,612);
\draw[BLACK, solid, line join=round, line cap=round, line width=1, fill=BLACK]
	(740,612) -- (731,618) -- (729,611) -- (740,612) -- cycle;
\draw[BLACK, solid, line join=round, line cap=round, line width=1]
	(370,594) -- (530,656);
\draw[BLACK, solid, line join=round, line cap=round, line width=1, fill=BLACK]
	(530,656) -- (519,656) -- (522,649) -- (530,656) -- cycle;
\draw[BLACK, solid, line join=round, line cap=round, line width=1]
	(530,650) -- (350,470);
\draw[BLACK, solid, line join=round, line cap=round, line width=1, fill=BLACK]
	(350,470) -- (359,475) -- (355,479) -- (350,470) -- cycle;
\draw[BLACK, solid, line join=round, line cap=round, line width=1]
	(346,470) -- (420,590) -- (530,654);
\draw[BLACK, solid, line join=round, line cap=round, line width=1, fill=BLACK]
	(530,654) -- (520,652) -- (523,646) -- (530,654) -- cycle;
\draw[BLACK, solid, line join=round, line cap=round, line width=1]
	(540,517) -- (670,473);
\draw[BLACK, solid, line join=round, line cap=round, line width=1, fill=BLACK]
	(670,473) -- (662,480) -- (659,473) -- (670,473) -- cycle;
\draw[BLACK, solid, line join=round, line cap=round, line width=1]
	(672,480) -- (640,520) -- (540,520);
\draw[BLACK, solid, line join=round, line cap=round, line width=1, fill=BLACK]
	(540,520) -- (550,517) -- (550,523) -- (540,520) -- cycle;
\draw[BLACK, solid, line join=round, line cap=round, line width=1]
	(875,460) -- (736,210);
\draw[BLACK, solid, line join=round, line cap=round, line width=1, fill=BLACK]
	(736,210) -- (744,217) -- (738,220) -- (736,210) -- cycle;
\draw[BLACK, solid, line join=round, line cap=round, line width=1]
	(590,172) -- (720,198);
\draw[BLACK, solid, line join=round, line cap=round, line width=1, fill=BLACK]
	(720,198) -- (710,199) -- (711,193) -- (720,198) -- cycle;
\draw[BLACK, solid, line join=round, line cap=round, line width=1]
	(720,198) -- (590,172);
\draw[BLACK, solid, line join=round, line cap=round, line width=1, fill=BLACK]
	(590,172) -- (600,171) -- (599,177) -- (590,172) -- cycle;
\draw[BLACK, solid, line join=round, line cap=round, line width=1]
	(720,200) -- (210,200) -- (210,580);
\draw[BLACK, solid, line join=round, line cap=round, line width=1, fill=BLACK]
	(210,580) -- (207,570) -- (213,570) -- (210,580) -- cycle;
\draw[BLACK, solid, line join=round, line cap=round, line width=1]
	(519,263) -- (474,180);
\draw[BLACK, solid, line join=round, line cap=round, line width=1, fill=BLACK]
	(474,180) -- (482,187) -- (476,190) -- (474,180) -- cycle;
\draw[BLACK, solid, line join=round, line cap=round, line width=1]
	(575,180) -- (529,263);
\draw[BLACK, solid, line join=round, line cap=round, line width=1, fill=BLACK]
	(529,263) -- (531,253) -- (537,256) -- (529,263) -- cycle;
\end{tikzpicture}
  \caption{The PN model representing the process of maintaining a glucose 
  homeostasis by the liver. During feeding, when the levels of glucose and insulin
  are high, and glucagon is not present, the glycolysis occurs. During fasting,
  when the level of glucagon is high, the synthesis of glucose occurs.}
  \label{fig2}
\vspace*{-0.3cm}
\end{center} 
\end{figure}
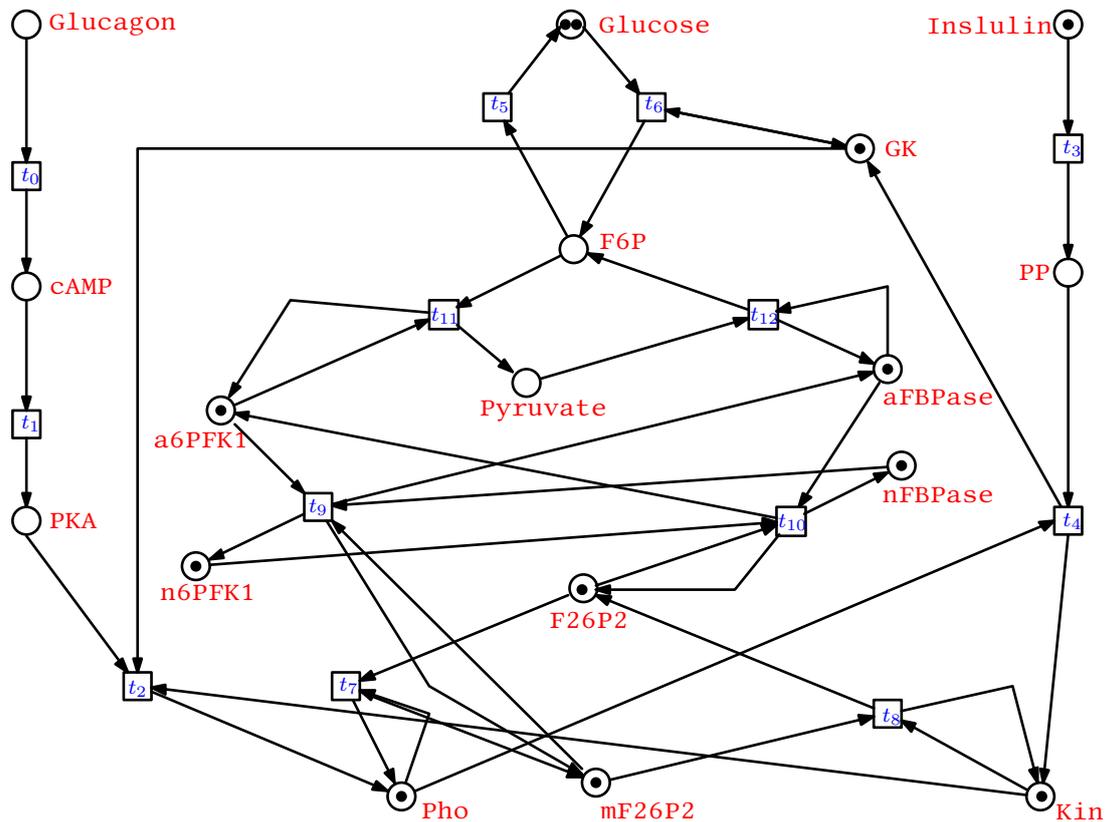

Notice, that markings 
of places \textit{Insulin}, \textit{Glucose} and \textit{Glucagon}
have the greatest impact on the dynamics of the model.
Markings of other places, associated with different enzymes
and substances like 6PFK2/FBPase2 (places \textit{Kin} and 
\textit{Pho}), F26P2 (places \textit{F26P2}, \textit{nF26P2}), 6PFK1 (places 
\textit{a6PFK1}, \textit{n6PFK1}) and FBPase (places \textit{aFBPase},
\textit{nFBPase}), would be established according to the presence of 
insulin and glucose or glucagon. 
This could happen because those places are included in different traps.
There are eight traps in the model:
$\{Kin, Pho\}$, $\{F26P21, mF26P2\}$, $\{a6PFK1, n6PFK1\}$, 
$\{aFBPase, nFBPase\}$, $\{n6PFK1, nFBPase\}$, $\{a6PFK1, aFBPase\}$,
$\{Pho, GK\}$, $\{Glucose, F6P, Pyruvate\}$. 
It means that if at least one of the place
from a trap is marked, the token would not leave the 
trap.
Hence, if 6PFK1 is less active  -- we observe less tokens in place \textit{a6PFK1}, and
at the same time more tokens in place \textit{n6PFK1}.
This behaviour is consistent with the actual biological processes.
The same holds for each trap. 
Transitions, which transfer tokens among 
places included in traps are affected by markings of places 
\textit{Insulin}, \textit{Glucose} and \textit{Glucagon}. Those markings in our analysis are fixed in the initial marking.
Nevertheless, in the more extended model of glucose regulation (which is our final goal)
markings of places \textit{Insulin}, \textit{Glucose} and \textit{Glucagon}
would arise from activity of particular parts of the model.
Especially, in the case of diabetics, by endogenous (produced in the pancreas) and exogenous (provided from the outside) insulin supply.
In the current state of the model those marking are administrated by hand.

The reachability graph of the net depicted in Figure~\ref{fig2}
(created using~\cite{snoopy}) contains 162 states with 702 arcs. 
The graph is obviously too big to be depicted properly, not to mention the detailed analysis. 
Even if we put only one token in the glucose place, the reachability graph of such a net is huge and therefore difficult to analyse (see Figure~\ref{fig.rg}).

\begin{figure}[ht]
\begin{center}
  \includegraphics[scale=0.4]{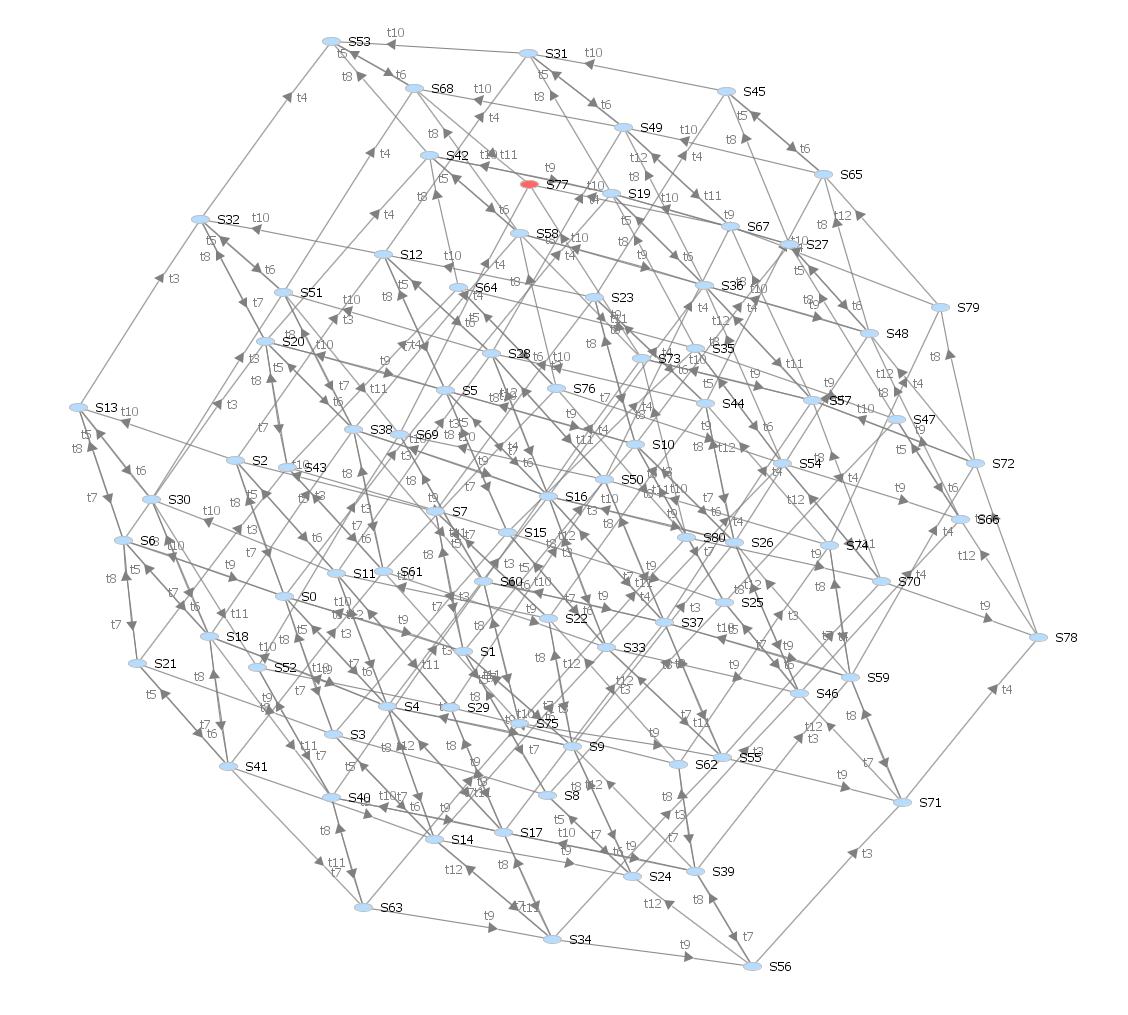}
  \caption{The reachability graph obtained for the PN model with the initial marking presented in Figure~\ref{fig2} with a slight change: place \textit{Glucose} contains 1 token. 
  Created using~\cite{snoopy}.}
  \label{fig.rg}
  \vspace*{-0.4cm}
\end{center} 
\end{figure}
For that reason, 
from now on, to study the behaviour of our model, 
we use reachability graphs obtained according to stubborn reduction (described in Preliminaries), computed by~\cite{charlie}. 
Note, that such graphs' properties are sufficient for the analysis of our model, making it easier and more transparent.

As mentioned above, the PN model represents the two basic processes of 
glucose homeostasis in the liver: glycolysis and synthesis of glucose. 
Those processed in a human body occurs alternately, one after the other. 
Hence, two kinds of initial markings of the PN model are possible. 
The first -- associated with the feeding state, and the second -- with the fasting.
We may start our analysis from any of those two. 
Let us begin with the model in the feeding phase. 
This results in the presence of glucose and insulin, and consequently,
in the initial marking, 
places associated with those substances are marked. In a human body,
in such a situation, the level of glucagon is very low (in comparison to glucose and insuline)
and in the model it is simply omitted. As mentioned above, markings of other places
are set according to the presence of glucose and insulin, we just have to make sure
that at least one place of each trap is marked. 
\newpage
The exemplary initial marking corresponding to the feeding phase is presented in Figure~\ref{fig2}.
For that marking, the stubborn reduced reachability graph is generated and depicted
in Figure~\ref{fig3}. One can easily notice, that every sequence of transitions executions
ends up in the (black) deadlock marking. 
Hence, we always start with tokens present in places
representing glucose and insulin, and eventually, those places are emptying, and the place
representing pyruvate becomes marked. It indicates that the glycolysis happened.
Obviously, someone might notice, that the presence of one or two tokens is not representative
for the modelling of biological processes.
Usually, in biological models, the 
number or value of tokens existing in a place, corresponds to the level of
the substance associated to the given place. 
The marking depicted in Figure~\ref{fig2} is choosen only for the illustration and to obtain a relatively small
reachability graph. We have performed also analysis with the larger numbers of tokens in
initial markings. The behaviour of the model stays the very same like in the case of the marking in Figure~\ref{fig2} -- each execution results in a deadlock marking, where places \textit{Insulin} and \textit{Glucose} 
are empty and place \textit{Pyruvate} contains the same number of tokens as place \textit{Glucose} 
in the initial marking. It indicates that the process of glycolysis has been accomplished.
Since the reachability graphs for Petri nets with certainly larger initial markings are too large to be shown and analysed here, we present the plot 
depicting the changes of the markings of the crucial places - Figure~\ref{fig3}.

\begin{figure}[ht]
\begin{center}
  \includegraphics[scale=1.1]{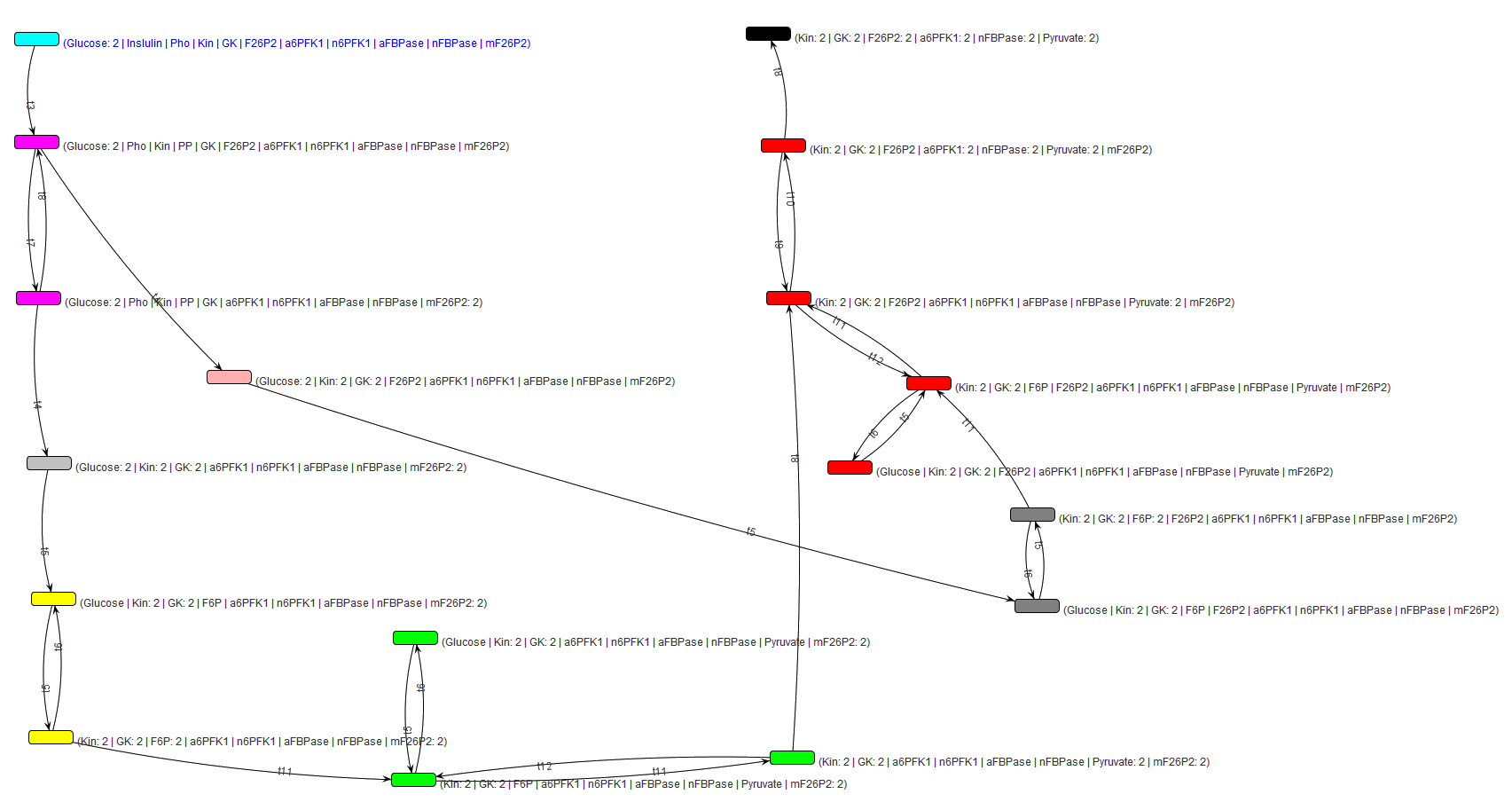}
  \caption{The reachability graph obtained for the PN model with the initial marking presented in Figure~\ref{fig2}. The initial marking is located top left and depicted blue, while the final (deadlock) marking is on the right corner of the figure (depicted black).
  Created using~\cite{charlie}.}
  \label{fig3}
\end{center} 
\end{figure}
After the feeding is finished, the fasting state starts, which means, that
the food from the environment is (temporarily) not more provided,
and the level of glucose in the blood is declining. In that condition of the organism, 
the pancreas is not producing insulin. When the level of glucose is low, the body is forced to use 
its own stored substances to ensure its suitable level in the blood. 
To obtain that goal, the pancreas starts to produce glucagon, and other organs, especially
the liver, react to the presence of glucagon. 
This situation is represented in our PN model as the 
deadlock marking obtained after the feeding, with additional tokens added to place \textit{Glucagon}.
As the result, the initial marking of the PN model to represent the fasting phase
would consist of: tokens in place \textit{Glucagon} and tokens in place \textit{Pyruvate},
empty place \textit{Glucose} and empty place \textit{Insulin} -- during fasting
the level of insulin would be much lower than the level of glucagon and in the model
it is omitted for simplification. Markings of other places are not that crucial as long as
each trap contains a token. The changes of number of tokens in places \textit{Glucose}, \textit{Pyruvate}
and \textit{Glucagon} are depicted in Figure~\ref{fig4}.
\begin{figure}[ht]
\begin{center}
  \includegraphics[scale=0.21]{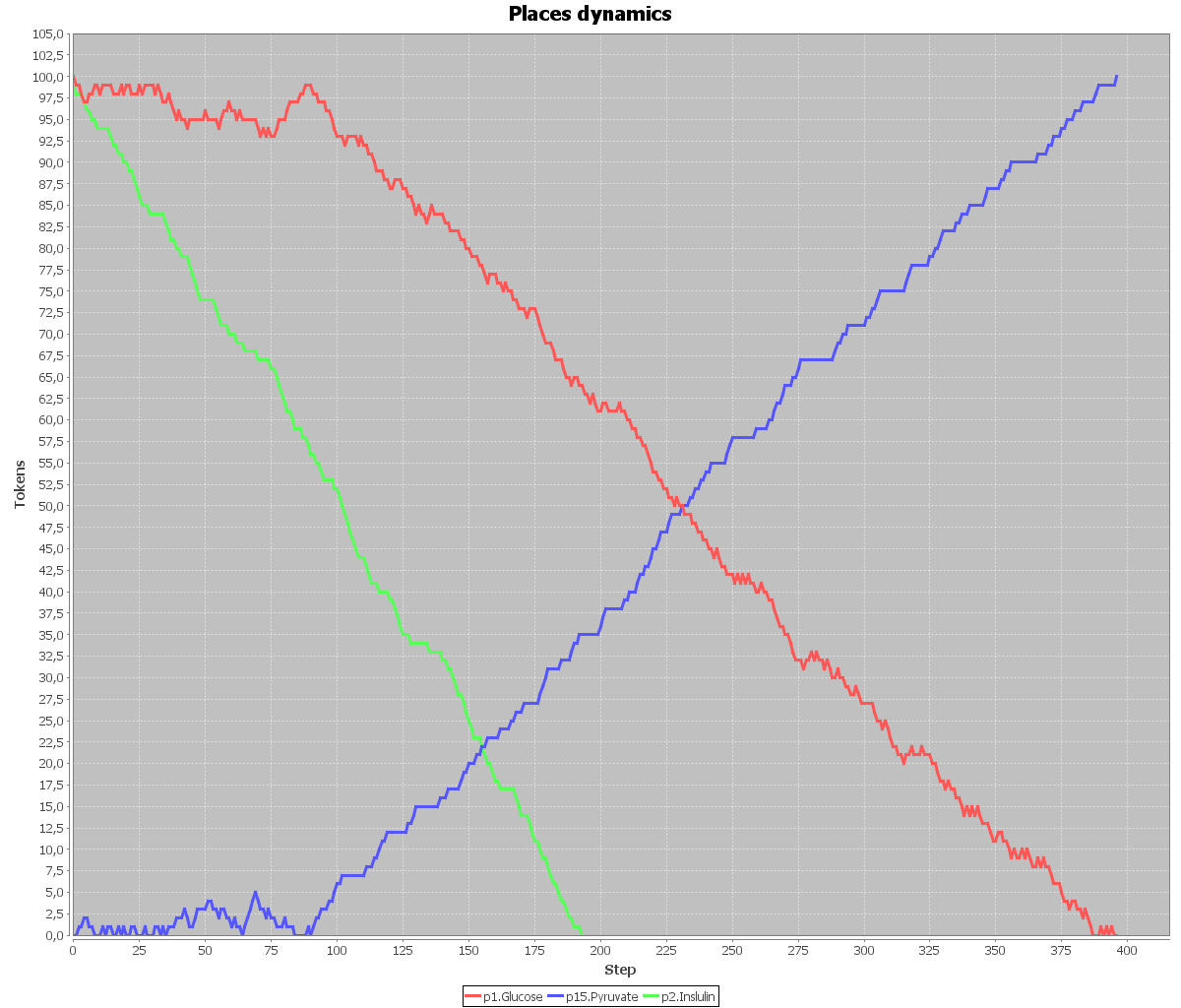}
  \caption{Changes in markings of places \textit{Glucose} (red), \textit{Pyruvate} (blue) 
  and \textit{Insulin} (green) in the feeding phase. In the initial marking places \textit{Glucose} and \textit{Insulin}
  contained each 100 tokens. At the end of places dynamics simulation place \textit{Glucose}
  contains zero tokens, and \textit{Pyruvate} contains 100 tokens.
  Created using~\cite{holmes2,holmes1}.}
  \label{fig4}
\end{center} 
\end{figure}
Figure~\ref{fig5} depicts the reachability graph of the PN model 
presented in Figure~\ref{netgraph}, but with a different initial marking determined as follows: \
places are marked as in the deadlock marking from Figure~\ref{fig3}
with two additional tokens in place \textit{Glucagon}. 
\begin{figure}[ht]
\begin{center}
  \includegraphics[scale=1.1]{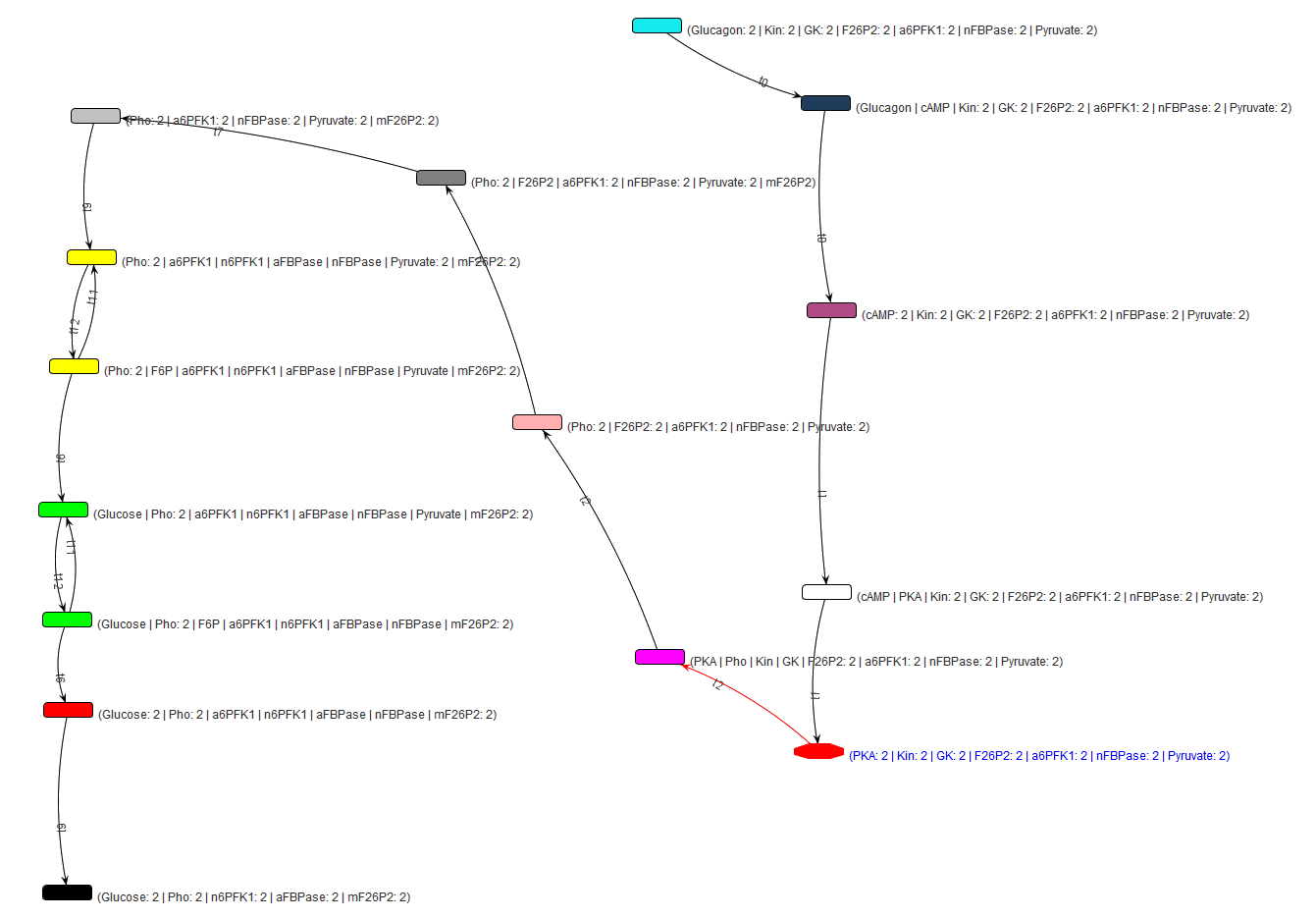}
  \caption{The reachability graph calculated for the PN model with the initial marking obtained
  from the deadlock marking, which is the final marking for the net in Figure~\ref{fig2}
  and two tokens added to place \textit{Glucagon}. The initial marking is located top right
   and depicted blue, while the final (deadlock) marking is on the left bottom corner 
   of the figure (depicted black). Created using~\cite{charlie}.}
  \label{fig5}
\vspace*{-0.3cm}
\end{center} 
\end{figure}
Like in the previous case, each execution results in the deadlock marking. In the initial marking, places \textit{Glucagon} 
and \textit{Pyruvate} are marked, in the deadlock marking, places \textit{Glucagon} 
and \textit{Pyruvate} are empty and place \textit{Glucose} contains the same number
of tokens as \textit{Pyruvate} in the initial marking. It corresponds to the process 
of glucose synthesis. 

Like above, we have also analysed the dynamic of the PN model
with larger number of tokens in the initial marking. 

The changes of number of tokens in places \textit{Glucose}, \textit{Pyruvate}
and \textit{Glucagon} are depicted in Figure~\ref{fig6}. It is easy to observe
that the number of tokens representing the level of glucose is increasing, 
while the number of tokens representing the level of pyruvate is deceasing. 
At the end of dynamics simulation, eventually there are as many tokens
in~\textit{Glucose} as there was in \textit{Pyruvate} in the initial state. 
The behaviour 
of the PN model with larger numbers of tokens is consistent 
with the behaviour of the model with the smaller one.
The deadlock marking obtained after the fasting phase
can be afterwards used as the 
initial marking for the simulation of the feeding and the next cycle 
of feeding and fasting can start.
\begin{figure}[ht]
\begin{center}
  \includegraphics[scale=0.21]{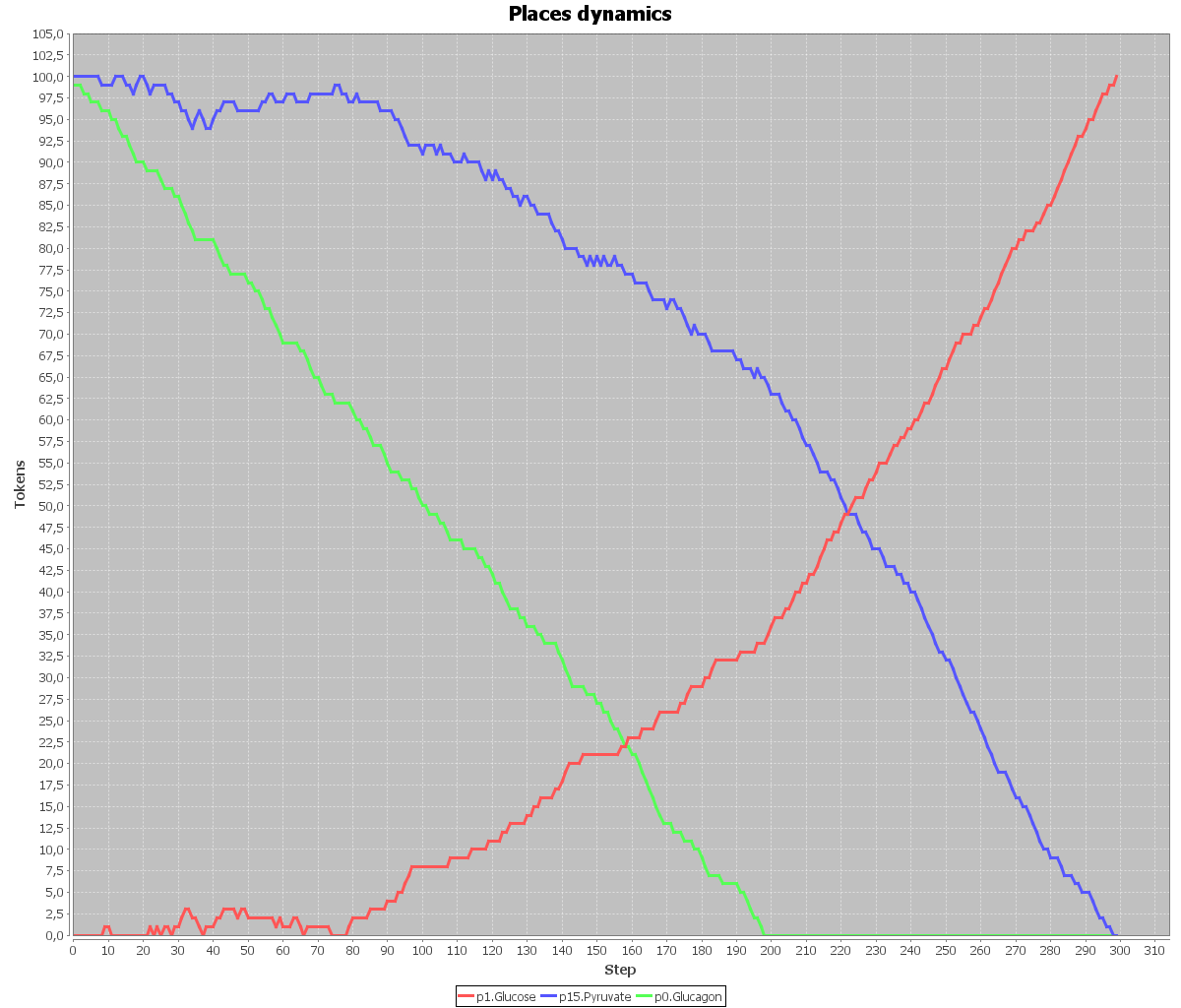}
  \caption{Changes in markings of places \textit{Glucose} (red), \textit{Pyruvate} (blue) 
  and \textit{Glucagon} (green) in the fasting phase. In the initial marking places \textit{Pyruvate} and \textit{Glucagon}
  contained 100 tokens each. At the end of places dynamics simulation place \textit{Pyruvate}
  contains zero tokens, and \textit{Glucose} contains 100 tokens.
  Created using~\cite{holmes1}.}
  \label{fig6}
\vspace*{-0.3cm}
\end{center} 
\end{figure}

\section{Conclusions and Future Work}

The paper constitutes the first step towards designing a model visualising the regulation of sugar levels in the body, which would aim to better understand the processes occurring in the body of a~healthy person, as well as a person suffering from diabetes.
As a first step, we model (with the use of Petri Nets) glucose regulating processes
in the liver. Two basic processes have been included in our model:
glycolysis during the feeding phase and synthesis of glucose
during the fasting state. The model preserves the
interactions between those processes, as shown in Section~\ref{model}.

We are aware that we are setting an ambitious goal and that subsequent steps towards it may require the use of more advanced tools, such as extensions to Petri nets (for instance: 
stochastic Petri nets, continuous Petri nets, 
fuzzy Petri nets, see \cite{marsan,fraca,cardoso}).
In the future, we also plan to verify our model 
(process modelling) and use it to find regularities and irregularities in the functioning of an unhealthy body.
We hope that the consequence of our actions will be greater awareness of diabetes and improvement of the physical and mental condition of people suffering from it.


\newpage
\begin{thebibliography}{99}

\bibitem{CGM2}
Battelino T., Alexander C.M., Amiel S.A., Arreaza-Rubin G., Beck R.W. , Bergenstal R.M., Buckingham B.A., Carroll J., Ceriello A, Chow E., Choudhary P., Close K., Danne T., Dutta S., Gabbay R., Garg S., Heverly J., Hirsch I.B., Kader t., Kenney J., Kovatchev B., Laffel L., Maahs D., Mathieu C., Mauricio D., Nimri R., Nishimura R., Scharf M., Del Prato S., Renard E., Rosenstock J., Saboo B., Ueki K., Umpierrez G.E., Weinzimer S.A., Phillip M.:
Continuous glucose monitoring and metrics for clinical trials: an international consensus statement,
The Lancet Diabetes \& Endocrinology,
2023.

\bibitem{cardoso}
Cardoso J, Valette R., Dubois D.:
Fuzzy Petri Nets: An Overview,
IFAC Proceedings Volumes, 
1996.

\bibitem{petri1}
Desel J, Reisig W.:
Place or Transition Petri Nets,
Lectures on Petri Nets, Vol. I: Basic Models, Advances in Petri Nets, 1491, 
1998.

\bibitem{fraca}
Fraca E., Haddad S:
Complexity Analysis of Continuous Petri Nets,
Application and Theory of Petri Nets and Concurrency,
Lecture Notes in Computer Science, vol 7927,
2013

\bibitem{heiner}
Heiner M., Schwarick M., Wegener J.-T.:
Charlie - An Extensible Petri Net Analysis Tool, Petri Nets 2015,
2015

\bibitem{NN3}
Jin X., Cai A., Xu T., Zhang X.: 
Artificial intelligence biosensors for continuous glucose monitoring, 
Interdisciplinary Materials, 
2023.

\bibitem{CGM1}
Lee I.,  Probst D., Klonoff D., Sode K.:
Continuous glucose monitoring systems - Current status and future perspectives of the flagship technologies in biosensor research,
Biosensors and Bioelectronics,
2021.

\bibitem{trap}
Liu GY., Barkaoui K.:
A survey of siphons in Petri nets. 
Information Sciences 363.
2016.

\bibitem{marsan}
Marsan, M.A.: 
Stochastic Petri nets: An elementary introduction,
Lecture Notes in Computer Science, vol 424,
1990.

\bibitem{Murata}
Murata T.:
Petri nets: Properties, analysis and applications,
Proceedings of the IEEE, vol. 77, no. 4,
1989.

\bibitem{NN2}
Nitesh P., Geeta R., Vijaypal S. D., Ramesh C. P.:
Diabetes prediction using artificial neural network,
Deep Learning Techniques for Biomedical and Health Informatics, Academic Press, 
2020.

\bibitem{NN1}
Prakash E. P., Srihari K., S. Karthik, Kamal M. V., Dileep P., Bharath Reddy S., Mukunthan M. A., Somasundaram K., Jaikumar R., Gayathri N., Kibebe Sahile:
Implementation of Artificial Neural Network to Predict Diabetes with High-Quality Health System. 
Computational Intelligence and Neuroscience, vol. 2022, Article ID 1174173, 2022.

\bibitem{holmes2}
Radom M. et al.:
Holmes: A graphical tool for development, simulation and analysis of Petri net based models of complex biological systems, Bioinformatics 33.23, 
2017.

\bibitem{raven}
Raven, P. H., G. B. Johnson, K. A. Mason, J. B. Losos, and S. R. Singer. "How cells harvest energy." In Biology. 10th ed. AP ed. (New York, NY: McGraw-Hill, 2014), 129.
Raven P. H.; Johnson G. B.; Mason K. A.; Losos J. B.; Singer S. R.:
How cells harvest energy,
In Biology 10th ed.,
2014.

\bibitem{Reisig}
Reisig W:
Petri Nets. An Introduction,
Part of the book series: Monographs in Theoretical Computer Science,
1985.

\bibitem{petri2}
Starke P:
Petri-{N}etze,
VEB Deutscher Verlag der Wissenschaften,
1980.

\bibitem{valmari1}
Valmari A.:
Error detection by reduced reachability graph generation. Proceedings of the 9th European Workshop on Application and Theory of Petri Nets,
1988.

\bibitem{valmari2}
Valmari A.:
A Stubborn Attack on State Explosion,
Formal Methods in System Design,
1992.

\bibitem{valmari3}
Valmari A., Hansen H.:
Stubborn Set Intuition Explained,
Transactions on Petri Nets and Other Models of Concurrency XII., vol 10470,
2017.

\bibitem{PAPER}
Xin G., Honggui L., Hang X., Shihlung W., Hui D., Fuer L., Alex J. L., Chaodong W.:
Glycolysis in the control of blood glucose homeostasis,
Acta Pharmaceutica Sinica B, Volume 2, Issue 4, 2012.\\
---
\
\bibitem{charlie}
Charlie, https://www-dssz.informatik.tu-cottbus.de/DSSZ/Software/Charlie

\bibitem{holmes1}
Holmes, http://www.cs.put.poznan.pl/mradom/Holmes/index.html

\bibitem{pipe2}
Pipe2, https://pipe2.sourceforge.net/

\bibitem{snoopy}
Snoopy, https://www-dssz.informatik.tu-cottbus.de/DSSZ/Software/Snoopy




---
\
\bibitem{loop4}
AAPS,
https://androidaps.readthedocs.io/en/latest/

\bibitem{loop3}
CamAPS FX,
https://camdiab.com/

\bibitem{DA}
Diabetes Atlas, 
https://diabetesatlas.org/

\bibitem{IDF} 
International Diabetes Federation,
https://federation idf.org

\bibitem{loop5}
Loop,
https://loopkit.github.io/loopdocs/

\bibitem{loop1}
MiniMed 780G System,
https://www.medtronic-diabetes.com/en-gb/minimed-780g-system-info

\bibitem{loop2}
Tandem Tslim Control IQ
https://www.tandemdiabetes.com/en-gb/home


\end{thebibliography}
\end{document}